\documentstyle[preprint,aps,amssymb,epsf]{revtex}  % PREPRINT format REVTEX 3.0
\tightenlines

\begin{document}
\draft                           % this command makes pacs numbers print

%-----------------------------------------------------------------------------

%\title{\vskip -0.5in\hfill\hfil{\rm\normalsize Printed on \today}\vskip 0.4in
\title{Quantum Computing: A View from the Enemy Camp}                           

\author{M.I. Dyakonov}  

\address{Laboratoire de Physique Math\'ematique,  Universit\'e 
         Montpellier 2, place E. Bataillon, 34095 Montpellier, France}

%\date{Received \quad}

\maketitle

%-----------------------------------------------------------------------------

\begin{abstract}

Quantum computing relies on processing information within a quantum system with many continuous degrees of freedom. The practical implementation of this idea requires complete control over all of the $2^n$ independent amplitudes of a many-particle wavefunction, where $n>1000$.  The principles of quantum computing are discussed from the practical point of view with the conclusion that no working device will be built in the forseeable future. 

\end{abstract}

%-----
---------------------------------------------------------------

\vskip 2cm

{\bf 1. Introduction}\\

We are witnessing an overwhelming rush towards "quantum computing".  Quantum information research centers are opening all over the globe, funds are generously distributed, and breathtaking perspectives are presented to the layman by enthusiastic scientists and journalists.  Many people feel obliged to justify whatever research they are doing by claiming that it has some relevance to quantum computing.  The impression is created that quantum computing is going to be the next technological revolution of the 21st century.
  
The comments below reflect my personal frustration with this state of affairs.  I will attempt to dispel the general enthusiasm and put the subject into a proper perspective.  No references to original work will be given. The reader may consult the excellent review by Steane\cite{Steane} on quantum computing with an historical survey of the development of the basic ideas and references (see also Ref.\cite{Book}). The skeptical remarks due to Landauer\cite{Landauer} and to Haroche and Raimond\cite{Haroche}  are also recommended.  The so-called quantum cryptography is beyond the scope of this presentation.

Quantum information theory is a respectable branch of mathematics, within which outstanding results, such as the famous Shor's algorithm and error correction codes, were obtained during the last decade.  Still, quoting Steane\cite{Steane} who has contributed much to the field, "The quantum computer is first and foremost a machine, which is a theoretical construct, like a thought experiment, whose purpose is to allow quantum information processing to be formally analyzed".

The natural questions are, first, do we need a real quantum computer,  and, second, can we build a practically working quantum computer in a foreseeable future? Although in my opinion the answer to both questions is negative, I will concentrate on the second one.  I will argue that quantum computing relies on storing and processing information within a physical system with a large number of continuous degrees of freedom, and that the requirements for the precision with which this physical system should be manipulated are absolutely unrealistic.
  
In fact, a quantum computer is a complex system of $n=10^3-10^5$ particles with two states each. While theoretically it may be quite enlightening to think about a complete description of the system by a grand wavefunction with its $2^n$ independent amplitudes, in practice it would be rather difficult to manipulate this wave function by controlled unitary transformations.
  
A classical example may help to better understand my point.  Take an ensemble of many oscillators, say a cubic centimeter of NaCl with its $10^{22}$ vibrational modes, or an array of LC circuits, and consider the following idea.  Let us store information by imposing prescribed amplitudes and phases for all of our oscillators.  We can then process this information by switching on external fields and interactions between oscillators.  This system would be a classical analog computer, and everyone can see immediately that such a proposal does not make sense for many obvious reasons.  Nonetheless, this classical system can very well mimic a quantum computer, although one would need $2^n$ classical oscillators to simulate a quantum computer with $n$ spins.  In spite of this fundamental difference, the basic idea of employing many continuous degrees of freedom for information processing is the same.  From the practical point of view, {\it this} property, not the quantum laws, makes all the difference with a digital computer.  It is by no means easier to manipulate the state of $10^5$ classical oscillators with the required speed and precision, virtually eliminating relaxation processes within this system, than to fulfill the same requirements for $10^5$ quantum spins.  In fact, the requirements for a quantum system are more stringent.
 
One of the main reasons for the interest in a quantum computer is  that it should be able to solve problems which are beyond the limits of classical computation.  The number of such known problems, like factorizing numbers greater than $10^{130}$, is currently between 2 and 10, depending on whether one counts similar problems as distinct, or not.  Certainly, the mere fact that such problems exist is of considerable theoretical interest.  However, from the practical point of view, their existence may be of interest to a relatively small minority involved in cracking cryptography codes.  Can this goal really justify years of efforts by an army of researchers? 
 
In the meantime, quite a lot of proposals for practical implementations of quantum computing have been advanced, mostly by theorists.  Some think that spins of electrons confined in quantum dots and interacting via the two-dimensional electron gas in the regime of the Quantum Hall Effect will do the job.  Others find it more natural to use anyons, hypothetical objects with a statistics intermediate between bosons and fermions that could exist in a two-dimensional space.  Less exotic ideas, like using conventional NMR techniques, were also put forward (see section 3 for a discussion of this idea).  Most of the ongoing experimental work is concerned with manipulating and measuring the states of 1-3 individual spins, or atoms.  Thus there is some hope (which I do not share) that within the next 20 years we will be able to factor the number 6, or even 15, by using Shor's algorithm!  It is for the reader to judge whether more exciting possibilities may exist.\\

{\bf 2.	What is a quantum computer and how does it work?}\\

The main idea is to replace the classical bit by the quantum bit, or qubit, a quantum two-level system, which can be thought of as the spin-up and spin-down states of a spin 1/2 particle.  These states are denoted as $|1>$ and $|0>$ respectively.  The general state of the spin is described by the wavefunction $\psi$, which is a superposition of the two states
$$\psi=a|0> + b|1>.	\eqno{(1)}$$

The complex amplitudes $a$ and $b$ satisfy the normalization condition $|a|^2 + |b|^2 = 1$, so that, apart from the irrelevant overall phase, $\psi$ is defined by two real parameters.  Classically speaking, these could be chosen as two angles describing the orientation of the spin vector.  More precisely, it is the orientation of the quantum mechanical average of the spin vector ${\bf S}$ (which would obey the classical Bloch equation in an external magnetic field).  The components of ${\bf S}$ are expressed through $a$ and $b$ in Eq. (1) as:
$$ S_{z} = (|b|^2 - |a|^2)/2, \hspace {0.3in} S_{x} = (ab^* + a^*b)/2,  \hspace {0.3in}  S_{y} = (ab^* - a^*b)/2i.	\eqno{(2)}$$ 

It is often stressed that a qubit can contain infinitely more information than a classical bit.  The reason is that the classical yes/no switch is replaced by an object with continuous degrees of freedom, the two polar angles determining the orientation of the spin vector.  This object certainly remains quantum, since one obtains the results given by Eq. 2 only in measurements averaged over an ensemble of spins, all of which are described by the wave function of Eq. (1).  An individual measurement of the spin component along any direction will always give values +1/2 or $-1/2$ with certain probabilities.  Thus, the probability of having a projection $\pm 1/2$ on the $x$-axis is $1/2 \pm S_{x}$.  

Quantum computing is based on controlled manipulation of the quantum state of $n$ spins, or $n$ qubits, where, if one wants to have a real computer, not just a demonstration toy, $n$ should be on the order of $10^3-10^5$ (the origin of this number will be explained below).  Let us begin with a two-spin or two-qubit system.  The general wavefunction is a superposition of four available states 
$$\psi= a|00> + b|01> + c|10> + d|11>  \eqno{(3)}$$
with four independent amplitudes, restricted only by the normalization condition. 

To get a feeling of what this state describes, we must again calculate the average values of the components of ${\bf S}_{1}$ and ${\bf S}_{2}$, but also correlations between the two spins, so that we will need average values of products $S_{1x}S_{2y}$, etc. The number of such independent physical quantities is exactly equal to the number of independent real parameters in Eq. 3.   Thus, Eq. 3 generally describes a state in which the two spins are correlated, the nature of correlation being defined by the values of amplitudes.
  
 Finally, a general state of $n$ spins is a superposition of $2^n$ basis states, like $|01001110...>$.  Each such state can be labeled by a single number $x$, which has the corresponding representation in the binary code.  Symbolically the wavefunction can be written as
$$\psi=\Sigma A_{x}|x>,  \eqno{(4)}$$
where the sum is over all values of $x$, from 0 to $2^n-1$, each $x$ represents a certain state of $n$ spins (example: $|5>$ stands for $|1011>$, if $n = 3$), and $A_{x}$ are the corresponding amplitudes.  In such a state, higher order correlations between all of the $n$ spins may exist.

Now we can have a definition of a quantum computer, which was basically put forward by Deutsch in 1985 (quoted from Stean \cite{Steane}).  A quantum computer is a set of $n$ qubits in which the following operations are experimentally feasible:\\

	1. Each qubit can be prepared in some known state $|0>$.

	2. Each qubit can be measured in the basis $\{ |0>, |1>\}$.

	3. A universal quantum gate (or set of gates) can be applied at will to any fixed-size subset of qubits.

	4. The qubits do not evolve other than via the above transformations.\\

To get just a glimpse of the beautiful ideas in quantum computing, consider the ingenious trick that is at the heart of quantum algorithms.  Suppose we have an integer variable $x$, between 0 and $2^n-1$, and a function $f(x)$, whose values are also integers in the same interval (for simplicity).  Assume that there exists an efficient (classical) algorithm for calculating $f(x)$ for a given $x$, however if $n$ is large, say 1000, calculation for all $x$ would take quite a lot of time.  The quantum computer, in a certain sense, performs all these calculations much faster, during a time which is not exponential, but polynomial in $n$. (I note in passing, that the difficulty of building a quantum computer increases exponentially with $n$, which is a kind of Nature's revenge).

Here is how it works.  Take two sets of $n$ qubits each.  One will serve to store the values of $x$, while the other is for the values of $f(x)$.  Start with the initial state $|0>|0>$, which means that all of the spins in both sets are in their down position.  Apply a unitary transformation to change the initial state to
$$\psi=2^{-n/2}\Sigma|x>|0>,	\eqno{(5)}$$
where again the sum is over all $x$, i.e. over all $2^n$ possible states of the first set. These states enter Eq. 5 with identical amplitudes, while the second set remains unchanged in all terms. 
 
Now apply another unitary transformation, $U_{f}$, which acts on any pure state $|x>|0>$ as
$$U_{f} |x>|0> = |x>|f(x)>, \eqno{(6)}$$
where, as before, $|f(x)>$ stands for the state of $n$ qubits of the second set, corresponding to the binary representation of the value of  $f(x)$.  Equation 6 means simply that $U_{f}$ calculates $f(x)$ for a given $x$.  However, when applied to the superposition in Eq. 5, this transformation gives
$$U_{f}\psi= 2^{-n/2} \Sigma|x>|f(x)>.	\eqno{(7)}$$

Thus we have obtained a state of the two sets of spins, which contains information on the values of $f(x)$ for all $x$.  This property is called quantum parallelism.  The important point is that in order to construct the transformation $U_{f}$, one does not need to pre-calculate the values of $f(x)$ for all $x$ (in which case the procedure would be senseless).  Instead, $U_{f}$  should contain the description of the algorithm to calculate $f(x)$, which is something much simpler.  It is the hardware that performs automatically all the $2^n$ applications of this (classical) algorithm.
  
If one measures the state of the system described by Eq. 7, one will obtain one of the pairs $(x, f(x))$ with equal probabilities.  So, this state is not useful by itself.  The idea of quantum algorithms (see Ref. 1 for details) is to perform further unitary transformations with the final goal to enhance the probability of the state, containing the answer to our problem.\\

\vskip 1cm
	
{\bf 3.	Discussion of feasibility}\\

Certainly, the ideas of quantum computing, which were only very briefly sketched in the previous section, are beautiful and appealing. However it should be fully realized that the practical implementation of these ideas requires complete control over all the intimate details of a many-particle quantum system.  I will now discuss the principles of a quantum computer, outlined above, from the practical point of view. 
 
Clearly, these mathematical postulates cannot be fulfilled exactly in any real experimental setup.  So, we must have an idea of the required precision and also of the number $n$ that is needed to outperform the classical computer.  If specially designed error-correction codes are used, an estimate for the tolerable noise level gives the order of magnitude of $10^{-5}$ per qubit per gate\cite{Steane}.  This estimate could be somewhat exaggerated, since only a simplified model of relaxation was used (uncorrelated errors in each qubit).  The values of $n$ of practical interest are defined by the maximal numbers that we should be able to store in order to beat the classical computer and also by the large quantity of additional qubits required for error correction (without error correction, a fantastic precision of $10^{-13}$ is required).  From these considerations an estimate $n=10^3-10^5$ is obtained \cite{Steane}.

While the basic points of the following discussion are independent on the physical nature of qubits, to be more specific, we will use the term "spins" instead of "qubits".  Thus, we must have some $10^3-10^5$ spins and the operation of our quantum computer depends on the possibility of the following:\\

1. We must be able to put them in the spin-down state with a precision $10^{-5}$.  No  problem about that: if these are electronic spins we can have this polarization  in a  magnetic field of 1T at temperatures below 100 mK.  However, this magnetic field will create some problems with postulate 4, to be discussed below.  (Here I assume that the precision required for the initial polarization, as well as for measurement of the system state, is of the same order as the estimated tolerable noise level.  The validity of this assumption is by no means evident, but apparently nobody has considered this problem theoretically.  Can the quantum computer tolerate an initial state, in which one of its $10^5$ spins points in the wrong direction, or a mistake in measuring one of the spins?  It should be noted that while physically the spins are equivalent, from the computational point of view it is not so: a mistake in the first spin changes the corresponding number $x$ by $2^{n-1}$, while a mistake in the $n$th spin changes it only by 1.  This difference does not matter if we do everything exactly, but it seems that it should be taken into account when one is concerned with the effects of errors.)\\

2. We must be able to measure the state of each spin, i.e. its projection on the direction of the magnetic field, presumably with the same precision.  This formidable task may be compared with measuring simultaneously all the velocities of a thousand atoms in a gas cell.  It is very difficult, though not impossible, to measure the state of a single spin.  This cannot be done directly, but rather through the influence of our spin on some process, such as photon emission (in this case we must measure the polarization of a single photon with a precision of $10^{-5}$, which is unheard of), or transport properties, which involve complex physics and additional interactions.  Any measurement scheme will be based on spin-orbital or exchange interaction with the environment, which will make it more difficult to satisfy postulate 4.  However, the problem of measuring one spin is infinitesimal compared to the problem of measuring simultaneously $10^5$ or even $10^2$ spins (remember, with a precision $10^{-5}$!).\\

3. We must have a universal quantum gate, that is, a device that can transform an arbitrary state of our system into another state of our choice via a unitary transformations of the system wave function.\\  

{\small This universal quantum gate is reminiscent of an old Russian joke.  The story goes that during World War II an inventor appeared with an idea of extreme military value.  Since at the time quantum cryptography did not yet exist, the inventor insisted that they take him to the very top, that is, to Stalin.
 
	- So, tell me what is it about.

	- It's simple, comrade Stalin. You will have three buttons on your desk, a green one, a blue one, and a white one.  If you press the green button, all the enemy ground forces will be destroyed.  If you push the blue button, the enemy navy will be destroyed.  If you push the white button, the enemy air force will be destroyed.
 
	- OK, sounds nice, but how will it work?

        - Well, it's up to your engineers to figure it out!  I just give you the idea.

Quantum-mechanically, anything that may happen in this world is a unitary transformation.  Thus the wave function of the destroyed ground force is related to that of intact ground force by a unitary transformation, and the same is true for the navy and the air force.  The required apparatus is illustrated in Fig. 1.  Before dismissing this analogy, the reader is invited to consider the difficulty of constructing a classical equivalent of the universal quantum gate, which must be easier.  Is there any hope of ever learning to transform an arbitrary state of $10^5$ classical oscillators into another prescribed state with a precision $10^{-5}$?}\\

%______________________ Fig.1___________________________________
\begin{figure}

\hskip 45pt \epsfxsize=350pt{\epsffile{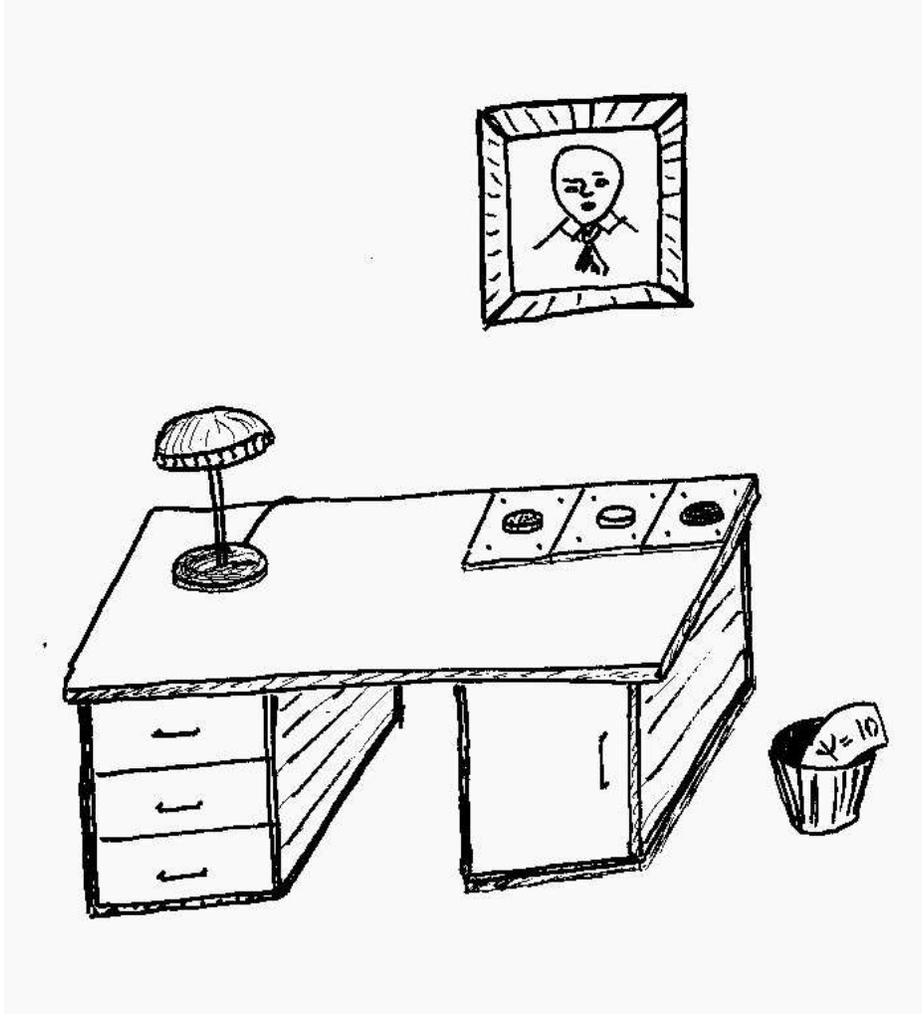}}
 
\vskip 15pt
 
\caption{A schematic view of Supreme Commander's workdesk after the installation of the universal military transformation device.}
 
\end{figure}
%________________________________________________________________

It was proved that an arbitrary transformation could be decomposed into a sequence of elementary transformations that influence either individual spins, or all possible pairs of spins.  The total number of such operations needed to beat the classical computer in factorizing numbers is estimated as $10^{10}$.  For each transformation, such as $U_{f}$, a special Hamiltonian of our spin system must be designed and implemented.  This means that magnetic fields should be applied individually and independently to each spin, but also $n(n-1)/2$ interactions between any two spins should be switched on and off at will.
 
There are some quite trivial difficulties.  An individual spin can be manipulated by applying either static magnetic fields or pulses of high frequency fields under the conditions of spin resonance.  Each spin must have an individual set of coils producing such fields, and each spin must be shielded from fields acting on other spins with enormous precision, which means that the inter-spin distance should be large enough (there should also be some space for the coils).  Then how can we make them interact?  If the field of 1 T, that is needed to provide the initial polarization, is present all the time, so that we can use spin resonance techniques, there are other problems.  It is easy to reverse the spin, by applying a $180^o$-pulse, however it is virtually impossible to put the spin in a fixed direction in the $xy$-plane, because of the continuous fast precession around the direction of the magnetic field.  (This point will be explained in other terms below).

This problem makes it impossible to achieve even the first step of quantum algorithms, the transformation, leading to the wave function in Eq. 5.  It can be easily seen that this wave function is the joint eigenfunction of all operators $S_{1x}, S_{2x}, ... S_{nx}$.  In other words, Eq. 5 describes a state in which all the spins point in the $x$-direction, while in the initial state they pointed in the $z$-direction.\\
   
4. Any evolution of the system, apart from the transformations imposed by our quantum gates, should be avoided.  Relaxation, or "decoherence" as a source of errors, was studied in many works in the framework of a simplified model of uncorrelated stochastic noise (the effects of noise on different qubits, or on the same qubit at different times, are assumed to be uncorrelated).  New and profound ideas of quantum error-correction were proposed \cite{Steane},\cite{Book} with the resulting estimate of $10^{-5}$ for the required precision, as was used above.  There are trivial problems that were not treated.  Suppose that some of the magnetic fields applied to individual spins, or interactions between pairs of spins do not have exactly the values they should have, which will be certainly the case for any practical device. This discrepancy would mean that the errors {\it are} correlated: each time a certain spin, or a certain pair of spins, is manipulated, the same type of error will occur. 
 
The list of similar problems could be easily enlarged. However, there is a more fundamental reason why in practice the postulate 4 will be violated.  So long as the energies of states $|0>$ and $|1>$ are different, which is the case for two-level atoms or spins in an external magnetic field, the general superposition in Eq. 4 will not remain unchanged, even if there is no relaxation and no quantum gates are applied.  The reason is simple: each amplitude $A_{x}$ will acquire a time-dependent phase $k\Omega t$, where $\Omega = E/\hbar$, $E$ is the energy difference, $t$ is the time, and $k$ is the number of qubits in the $|1>$ state within a state $x$.  (For example, $k = 3$ for the state $|1011>$, corresponding to $n = 4, x = 11$).  This time-dependent phase difference is the quantum-mechanical way of saying that each spin performs a fast precession around the direction of the magnetic field.  Thus, our $2^n$ states will be divided into groups with the same phase (same values of $k$), which are physically similar, but computationally quite different (for example, states $|1000> (x=8)$ and $|0001> (x=1)$ fall into one group).
  
For atomic hyperfine levels, the frequency  $\Omega  = E/\hbar$ is typically on the order of $10^{10}- 10^{12}$ s$^{-1}$, and it is impossible to fix the relative phase of states with different $k$ on such a time scale.  As a result, the relative phases will have random values at the moments when our gates are applied, which will completely disorganize the performance of our quantum computer.   For electron spins in a magnetic field of 1 T, the corresponding frequency  $\Omega$  is on the order of $10^{11}$ s$^{-1}$.  Even if we move the fundamental frequency to an accessible range, say 100 MHz, by lowering the field to 0.001 T and making the temperature 0.1 mK, it still will not help.  There always will be higher harmonics with the difference in $k$ on the order of $n = 10^3-10^5$, whose coherence it would be impossible to preserve.  The existence of the higher harmonics with frequencies $p\Omega$ ($p = 1, ..., n$) has a simple physical explanation: they are the frequencies of evolution of the $p$-spin correlations of the type $<S_{1x}S_{2x} ... S_{px}>$. Since each spin makes a precession in the $xy$-plane with a frequency $\Omega$, such a product will have all the harmonics, up to $p\Omega$ .

It appears that the only way to avoid this fast and unwanted evolution of our wave function is to have qubits with degenerate or quasi-degenerate states.  This means that we will have to abandon the spin resonance method and switch off the initial field of 1 T.  This switching will take a long time, during which some relaxation could take place.  Also the switching-off process will, most probably, perturb our spin system.  Besides, we will have to do this switching on and off each time we need the state $|00000...>$ for the next step of calculations.  After we have made the spin states degenerate, we will have to use quasi-static magnetic fields to independently rotate individual spins in the desired direction with the desired precision (please, design the experimental setup).
  
It seems that at this stage it would be pointless to discuss numerous additional problems related to the possibilities of controlling interactions between any two spins, which is indispensable for quantum computing.

In summary, I do not believe that it will be possible to construct even a toy (but working!) quantum computer with three qubits.\\ 

\vskip 1cm

{\bf 4. The classical limit of a quantum computer}\\

There is no impenetrable wall between quantum and classical physics.  Quantum effects may gradually disappear when some parameter is changed.  Under certain conditions, even such small objects as electrons, are very well described by Newton's laws.  For a solid-state physicist, a familiar example is provided by magnetotransport phenomena.  At room temperature, transport is well described by the classical Drude theory.  As the temperature is lowered, small Shubnikov-de Haas oscillations appear, which are due to the quantization of the electron energy in magnetic field.  In the strong magnetic field limit one enters the extreme quantum Hall effect regime.  Note that the condition for a classical description to be valid is not the requirement that the Landau energy spacing be smaller than $\Delta E = \hbar/\tau$, where $\tau$ is the relaxation time.  Their ratio is a classical parameter $\omega_{c}\tau$, where $\omega_{c}$ is the cyclotron frequency.  This parameter, which enters the classical Drude formulas, may be quite large and still the classical description will hold.  The true criterion is the ratio between the Landau spacing and the thermal energy, $kT$.  The classical description fails when this ratio becomes large enough.

In view of these remarks, the following questions seem to be justified.  What is the classical limit of a quantum computer?  What are the conditions for achieving this limit?  These are difficult questions, which were never addressed properly by the quantum computing community.  Apparently, most people in the field believe that the classical limit of a quantum computer is a digital computer, employing switching between spin-up/spin-down states as a classical bit, and that this limit is achieved when the coherent superposition of states in Eq. (4) is destroyed by relaxation processes.  This is supposed to happen when the energy spacing between the spin-up and spin-down states, due to an applied magnetic field, becomes less than $\Delta E = \hbar/\tau$.  (Another way of expressing this idea is to say that the classical limit is achieved when the off-diagonal elements of the density matrix disappear). Just as in the example above, this statement is not correct.  The quantity $\Omega\tau$, where $\Omega$ is the spin precession frequency in the external magnetic field, is a classical parameter (it does not contain the Planck constant) describing the quality of the spin resonance, so that quantum versus classical behavior cannot be chosen depending on the value of this parameter.  One need not go to the over-damped situation ($\Omega \tau < 1$) to insure classical behavior. (In terms of the density matrix, it should be noted, first, that a matrix that is diagonal in one basis may become non-diagonal in another basis and,  secondly, that it is the non-diagonal density matrix that transforms into the classical distribution function in the classical limit, see, for example the Wigner density matrix.)

In my opinion, the classical limit of the quantum computer is a classical analog machine that stores information in the mutual orientations of a large number of classical magnetic moments (spins).  Evidently, the practical possibility of building such a machine is questionable, to say the least.  However it is much, much easier than to build a true quantum computer.

To shed more light on these questions, let us for a moment replace the two-state qubit ($S$ = 1/2) by a "many-state qubit" (high spin $S>>1$).  For example for $S$ = 9/2, which is the spin value for the $^{113}$In nucleus, one has $2S+1 = 10$ states.  This change will not affect the basic ideas for quantum computing, except that now we will have to abandon the binary code for writing numbers in favor (for example) of the familiar decimal code.  It is well known that for high quantum numbers the behavior of any quantum system approaches that of a corresponding classical system. Thus, under application of external fields, various interactions, and so on, the hardware of our high-spin quantum computer should behave very similarly to a classical system of spins.  However, under ideal conditions it should still remain a quantum computer. 
 
The paradox is resolved by taking into account the effects of finite temperature, finite precision of our unitary transformations, or finite accuracy of our measurements.  As soon as the precision becomes such that we cannot distinguish between neighboring states $m$ and $m+1$, where $m$ is the projection of ${\bf S}$ on a chosen axis (though we can still distinguish between states for which $|m_{1}-m_{2}| >> 1$), we will see purely classical behavior.  (This path is the usual one from quantum towards classical mechanics.  A tennis ball in a rectangular one-dimensional potential well obeys textbook quantum mechanics.  For high energy levels the wave function is a standing wave, but since the wavelength is smaller than the precision of any imaginable instrument, we will see the classical distribution function, which is uniform throughout the well.  Note again that the classical limit does not require that the energy spacing be smaller than $\hbar/\tau$, or that the tennis ball density matrix be diagonal.)  But at this point the quantum computer will stop working as such, since we will be making random mistakes of $\pm 1$ in each digit of all the numbers written in our $(2S+1)$-code.  Instead, we will have an analog machine consisting of classical spins.  

In this context, a remark concerning some proposals to use NMR techniques for quantum computing is in order.  At first glance, the advantage of nuclear spins is in their extremely long relaxation times, due to the very small value of nuclear magnetic moments and consequently to their poor coupling to external fluctuating magnetic fields.  However, for the same reason it is extremely difficult to obtain full nuclear polarization and it is practically impossible to measure the state of an individual nuclear spin.  In order to overcome these difficulties, it is suggested to use not an individual spin, but rather a cluster of, say, $10^{15}$ nuclear spins, which may be weakly polarized.  A thousand of such clusters are supposed to play the role of the qubits.
  
In fact, the idea is completely misconceived.  A weakly polarized cluster of many nuclear spins behaves like a classical spin, and cannot be represented by a wave function given by Eq. (1).  Accordingly, a measurement of the spin per nucleus will always give the small average value with a $100\%$ probability (more precisely, with an uncertainty $10^{-15/2}$).  The situation would not change even if the nuclear polarization were to reach $99.99\%$.  This is the reason one actually does not need quantum mechanics to understand most of the phenomena pertaining to NMR or, more generally, to the average spin of large ensembles.  Many clusters would work as a classical analog machine, described above, but by no means like a quantum computer.\\ 

\vskip 1cm	

{\bf 5.	Conclusions}\\

We should make a clear distinction between the quantum computer of the abstract information theory and the working quantum computer, between a thought experiment and a real experiment.  For a mathematician, rotating a vector in a huge Hilbert space is a routine algebraic operation, while doing it in practice may be possible or not, depending on the required precision, the dimension of the space, and other conditions imposed by the real physical world.

Modern ideas linking quantum mechanics and information theory are quite fascinating, providing better insight into both theories, but this advance is not related in any way to the possibility of building a working quantum computer.
 
The practical performance of a quantum computer is based on manipulating on a microscopic level and with an enormous precision a many-particle physical system with continuous degrees of freedom.  Obviously, for large enough systems, either quantum or classical, this becomes impossible, which is why such systems belong to the domain of statistical, not microscopic physics.  The question is whether a system of $n = 10^3-10^5$ quantum spins, needed to beat the classical computer in solving a limited special class of problems, is large enough in this sense.  Can we ever learn to control the $2^n$ amplitudes defining the quantum state of such a system?  Based on the previous analysis, my answer is {\it no, never}.

There is an intermediate link between a quantum and a digital computer, which is a classical analog machine, like a system of  $10^5$ oscillators or classical spins.  Why not try to build such a machine first, and then, when we see that it works, try to accomplish the next, orders-of-magnitude more difficult task of constructing a quantum computer with $10^5$ quantum spins?  Personally, I do not believe that even such a classical analog machine of sufficient complexity will ever work.

The fashion of quantum computer building will gradually die away, the sooner, the better.  Research in atomic and spin physics is interesting and useful on its own and need not be justified by irresponsible projects and promises.\\

\vskip 1cm

{\bf 6.	Acknowledgments}\\

I am indebted to Andrew Steane for illuminating correspondence.  I have also benefited from discussions with Sergei Meshkov, G\'erard Mennessier, and Serge Luryi.

%-----------------------------------------------------------------------------


\begin{references}
                               

\bibitem{Steane}

A. Steane, "Quantum computing," Reports Prog. Phys. 61, 117 (1998); LANL e-print quant-ph/9708022, http://xxx.lanl.gov.

\bibitem{Book}

"The physics of Quantum Information", D. Bouwmeester, A.Ekert, A.Zeilinger (editors), Springer (2000).

\bibitem{Landauer}
	
R. Landauer, "The physical nature of information," Phys. Lett. A 217, 188 (1996).

\bibitem{Haroche}

S. Haroche and J-M. Raimond, "Quantum computing: dream or nightmare?" Phys. Today, August (1996), p. 51.

\end{references}
\end{document}